\documentclass[12pt,preprint]{aastex}
\usepackage{natbib}
\def\idm#1{{\mbox{\scriptsize #1}}}

\shortauthors{K. Go\'zdziewski and A.~J. Maciejewski}
\begin{document}
\title{The Janus head of the HD~12661 planetary system}
\author{Krzysztof Go\'zdziewski\altaffilmark{1}, and 
        Andrzej J. Maciejewski \altaffilmark{2}}
\altaffiltext{1}{Toru\'n Centre for Astronomy, 
  N.~Copernicus University,
  Gagarina 11, 87-100 Toru\'n, Poland; k.gozdziewski@astri.uni.torun.pl}
\altaffiltext{3}{Institute of Astronomy,                                                       
  University of Zielona G\'ora,
  Podg\'orna 50, 65-246 Zielona G\'ora, Poland;
  maciejka@astro.ia.uz.zgora.pl}
\begin{abstract}
In this work we perform a global analysis of  the radial velocity curve of
the HD~12661 system.  Orbital fits that are obtained by the genetic and
gradient algorithms of minimization reveal the proximity of the system to the
6:1 mean motion resonance.  The orbits are locked in the secular resonance
with apsidal axes librating about $180^{\circ}$ with the full amplitude
$\simeq (90^{\circ}$,$180^{\circ})$. Our solution  incorporates the mutual
interaction between the companions.  The stability analysis with the MEGNO
fast indicator shows that the system is located in an extended stable zone of
quasi-periodic motions. These results are different from those
obtained on the basis of the orbital fit published by \cite{Fischer2003}.
\end{abstract}
\section{Introduction}
The recently discovered planetary systems, HD~12661 and HD~38529
\cite[]{Fischer2003}, are among about a dozen of multi-planetary systems.
Their dynamics  are the subject of extensive work carried out by many
researchers. The   knowledge of these systems changes rapidly, as more and
more observational data are gathered. Remarkably,  most of the new
multi-planetary systems have a resonant nature. The mean motion resonances
(MMR) and/or  the secular apsidal resonances (SAR) are most likely present in
$\upsilon$~Andr, HD~82943, Gliese~876, 47~UMa, 55~Cnc, and HD~12661
\cite[Table~8]{Fischer2003}. 

The orbital fit to the radial velocity (RV) observations of the HD~12661
system has been determined by \cite{Fischer2003}. They discovered  two giant
planets, b and c, with masses $m_{\idm{b}} \simeq 2.3 \mbox{M}_{\idm{J}}$,
and    $m_{\idm{c}} \simeq 1.57 \mbox{M}_{\idm{J}}$, revolving around the
parent star in elongated orbits, $e_{\idm{b}} \simeq 0.35$, $e_{\idm{c}}
\simeq 0.2$ and semi-major axes $a_{\idm{b}} \simeq 0.83$~au and $a_{\idm{c}}
\simeq 2.56$~au, respectively. These data permit to classify the HD~12661
system as a planetary hierarchical triple system. Recently,  using the  
initial conditions provided by \cite{Fischer2003}, \cite{Gozdziewski2003} and
\cite{Lee2003}, investigated the dynamics of the HD~12661 system.
\cite{Lee2003} studied coplanar dynamics in the framework of an analytical
secular theory. The  work of \cite{Gozdziewski2003}, mostly numerical, is
devoted to the stability analysis of the system, in a wide neighborhood of
the initial condition, also for non-coplanar systems. The results of these
papers are in accord. They  reveal that the HD~12661 system is close
to the 11:2 MMR, accompanied  by the SAR, with the critical angle
$\varpi_{c}-\varpi_{b}$ librating about $180^{\circ}$. This type of  SAR is
the first such case found among the known planetary systems \cite[]{Lee2003}.
The octupole-level secular theory of these authors explains that the 
planetary system is located in an extended resonance island in the phase
space, and it  helps to understand why the resonance persists in  wide ranges
of orbital parameters. This is also true for non-coplanar systems
 and wide ranges of  planetary masses
\cite[]{Gozdziewski2003}.

Our dynamical study of the HD~12661 system \cite[]{Gozdziewski2003} have been
based on the first 2-Keplerian fit published by the California and Carnegie
Planet Search Team on their WEB site\footnote{\tt http://exoplanets.org/}. 
Recently, an updated fit to the RV data appeared \cite[Table
2]{Fischer2003}.  A preliminary dynamical analysis of this initial condition
does not bring qualitative changes to the system dynamics.  However, the
results of remarkable papers by \cite{Laughlin2001} and \cite{Lee2003} 
inspired us to analyze the RV data again, in the framework of the full, nonlinear
$N$-body dynamics. Although the RV observations are most commonly modeled by additive
Keplerian signals, the mutual interaction between giant planets can introduce
substantial changes to such simple model. A further nuance flows from the
fact that the 2-Keplerian fits are mostly interpreted as osculating,
astrocentric elements. In fact, the fits should be interpreted as Keplerian
elements related to the {\em Jacobi} coordinates \cite[]{Lee2003}.  These
refinements seem to change qualitatively the view of the dynamical state of
the HD~12661 system, as it can be derived from the current RV data.
 
\section{Orbital fits}

We used  the updated RV observations of the HD~12661 system
\cite[Table 4]{Fischer2003}, containing 86 data points, collected in Keck and Lick
observatories.  The reported uncertainties are $\simeq 3$-$6$~ms$^{-1}$ for
the Keck data and $\simeq 7$-$17$~ms$^{-1}$ for the Lick observations.

In the first attempt to find the initial condition to the RV data we applied
the genetic algorithm (GA). To the best of our knowledge, this method of
minimization was used for analyzing the RV of $\upsilon$~Andr by
\cite{Butler1999} and \cite{Stepinski2000},  the Gliese~876 system by
\cite{Laughlin2001}, and  55~Cnc by \cite{Marcy2002}.  
Basically, the genetic scheme makes it possible to find the {\em global}
minimum of the $\chi^2$ function. Our experiments confirm remarks of
\cite{Stepinski2000} that the method is inefficient for finding very accurate
best-fit solutions, but it provides good starting points to fast and
precise gradient methods, like the Levenberg-Marquardt (LM) 
scheme \cite[]{Press1992}.  These two methods of minimization complement each
other. We used publicly available code PIKAIA v.~1.2\footnote{\tt
http://www.hao.ucar.edu/public/research/si/pikaia/pikaia.html} by
\cite{Charbonneau1995}, a great advocate of the genetic algorithms.

In the simplest case,  the minimized function is given by the following
formula
\[
 \chi^2 = \sum_{i=1}^N \left[
 \frac{ V_{\idm{k}}(t_i,{\bf p})+V_0-V(t_i)}{\sigma(t_i)} \right]^2,
\]
where $N$ is the number of observations   of the radial velocity $V(t_i)$ at
moments of time  $t_i$,  with uncertainties $\sigma(t_i)$, modeled by a sum
of Keplerian signals   $V_{\idm{k}}(t_i,{\bf p}) =V_{\idm{b}}(t_i,{\bf
p}_{\idm{b}})+V_{\idm{c}}({t_i,\bf p}_{\idm{c}})$, and the velocity  offset
$V_0$. The fit parameters, ${\bf p}=({\bf p}_{\idm{b}},{\bf p}_{\idm{c}})$,
where  ${\bf p}_{\idm{p}}=(K_{\idm{p}},n_{\idm{p}},e_{\idm{p}},
\omega_{\idm{p}},T_{\idm{p}})$, for every planet ${\mbox{p}}
={\mbox{b}},{\mbox{c}}$, are: the amplitude $K_{\idm{p}}$, the mean motion
$n_{\idm{p}}$, the eccentricity $e_{\idm{p}}$,  the argument of periastron
$\omega_{\idm{p}}$, and the time of periastron passage $T_{\idm{p}}$.  The
orbital elements emerging from these parameters should be related to the
Jacobi coordinates~\cite[]{Lee2003}. 
Recently, we drew a similar conclusion independently \cite[]{Gozdziewski2003a}. 
In our fit model, the planetary system's reference
frame is chosen in such  way that the $z$-axis is directed from the observer
to the system's barycenter and $V(t)$ is the $z$-coordinate of star velocity.
The relation between $K_{\idm{p}}$  and the mass of a planet $m_{\idm{p}}$,
the mass of the host star $m_*$  and the geometric elements,  related to the
Jacobi coordinates, is given by  $ K_{\idm{p}} =\sigma_{\idm{p}} a_{\idm{p}}
n_{\idm{p}} \sin i$,  where, for the three-body system,  $\sigma_{\idm{b}}=G
m_{\idm{b}}/(m_*+m_{\idm{b}})$ for the first planet, $\sigma_{\idm{c}}=G
m_{\idm{c}}/(m_*+m_{\idm{b}}+m_{\idm{c}})$ for the second companion.  For
efficiency reasons, the code for the dynamical analysis, incorporating the
MEGNO indicator, works internally in astrocentric coordinates, thus,  when it
is required, the Jacobi coordinates are transformed to these coordinates. In
this sense, the astrocentric elements are considered only as a formal
representation of the initial condition\footnote{Although it is quite
obvious, we would like to remark that the results of the MEGNO tests are
coordinate independent.}.  
Together with $\chi^2$, we use 
$
 \sqrt{\chi^2_{\nu}} = \sqrt{\chi^2/\nu},
$ 
where 
$\nu=N-N_p-1$ is the number of  degrees of freedom and
$N_p$ is the total number of  parameters.

The PIKAIA code is controlled by a set of 11 parameters. We set the
most  relevant  as follows: the population number to 256, the number of
generations to a rather high value 8000--12000, and the variable mutation rate in
between 0.005 to 0.07 with crossover probability equal to 0.95. The PIKAIA
runs, restarted many times, resulted in a number of qualitatively similar
fits, with $\sqrt{\chi^2_{\nu}} \simeq 1.44$ and  RMS~$\simeq 8.2$~ms$^{-1}$.
The RV data have been modeled by 2-Keplerian signal and   one RV offset. We
found that, surprisingly, all these solutions,  not only have a slightly
smaller $\sqrt{\chi^2_{\nu}}$ and RMS than reported by  \cite{Fischer2003}
for the 2-Keplerian model ($\sqrt{\chi^2_{\nu}} \simeq 1.46$ and  RMS~$\simeq
8.8$~ms$^{-1}$), but also differ from the former fit substantially. The
periods ratio is very close to 6:1, and not to 11:2!   
Also, the  initial $e_{\idm{c}} \simeq 0.1$ is smaller
than previous estimates.  

Actually, because the Doppler measurements have been performed in two
different observatories,  we should account for two independent  velocity
offsets $V_0$ and $V_1$, different for the the Lick
and Keck observations.  Data that  are inhomogeneous in this sense
were  analyzed by \cite{Stepinski2000} and \cite{Rivera2001}. Indeed,  the
GA  incorporating the RV model with two velocity offsets, 
makes it possible to obtain a significantly better solution
($\sqrt{\chi^2_{\nu}} \simeq 1.37$ and RMS~$\simeq 7.9$~ms$^{-1}$). 

We verified this fit by determining the quasi-global minimum of $\chi^2$ as a
function of $(P_{\idm{c}},e_{\idm{c}})$.  We searched for the best fit
solutions with these two parameters fixed. Because $P_{\idm{b}}$ is
determined very well,  its value, as well as $e_{\idm{b}}$, approximated by
the best GA solution for the 2-Keplerian model, was selected for the 
starting point in the LM method. Next, varying the initial phases  (the
arguments of periastron and the mean anomalies) of the two companions, with
the step $\simeq 30^{\circ}$, we calculated the best fit solution with the LM
algorithm. The results of this experiment  are given in Fig.~\ref{fig:fig1}.
This scan reveals that the best fit solution is localized in a flat minimum. 
Its parameters agree very well with the GA solution. This test confirmed 
that the GA fit is really global. Finally, we refined this fit by the LM
algorithm, driven by the full model of the dynamics, incorporating the mutual
gravitational  interaction between  the planets. The best GA fit was
used as a starting point to the gradient search.  The result  is shown in
Table~\ref{tab:fits},  and named  the LM1 fit. The synthetic RV curve is
shown in Fig.~\ref{fig:fig2}.

To estimate the errors of this solution we used, at the first attempt,
a method relying on a determination of the $\chi^2$ confidence
intervals \cite[]{Press1992}. We were warned by the referees, that
this method has drawbacks and the most relevant is that using it we do
not account for the stellar ``jitter''. Even for a chromospherically
inactive HD~12661 star, the uncertainty, $\sigma_{\idm{jitter}}$,
contributed by the jitter to the RV measurements, can be relatively
high. The referee, Gregory Laughlin, suggested to us  using the estimate
$\sigma_{\idm{jitter}} \simeq 5$~ms$^{-1}$, based on the data by
\cite{Saar1998}. Actually, the observations have to be weighted by
$\sigma = \sqrt{\sigma_{\idm{obs}}^2 + \sigma_{\idm{jitter}}^2}$,
where $\sigma_{\idm{obs}}$ are ``pure'' instrumental errors. Such joint
uncertainty, accounted for calculating $\sqrt{\chi^2_{\nu}}$, leads to
a value $\simeq 1$, thus giving a statistical indication of an
adequate model of the RV data. To estimate the parameters errors we
synthesized about $250$ sets of ``observations''. To every original RV
measurement we added a Gaussian noise with the mean dispersion
$\sigma_{\idm{K}}=3.4$~ms$^{-1}$, $\sigma_{\idm{L}}=7.6$~ms$^{-1}$ for
observations gathered by Keck and Lick spectrometers, respectively,
and the Gaussian noise of the stellar jitter, with the dispersion
$\sigma_{\idm{jitter}}=5$~ms$^{-1}$. Next, the Newtonian model of
dynamics was fitted to every such synthetic data set with the LM
algorithm which started from the LM1 solution. The mean values of the
fit parameters are given in Table~\ref{tab:fits}, and called the LM2
fit. Finally, the dispersions of these orbital parameters are adopted
as the mean uncertainties of both the LM1 and LM2 fits. In
fact, both solutions are the same, with respect to the error bounds.

\section{Stability analysis}

To investigate the dynamical stability of the LM1 and LM2 fits we used the
fast indicator called MEGNO. This technique was advocated by us in a
series of recent papers, see, e.g,
~\cite[]{Gozdziewski2001a,Gozdziewski2003}.
The MEGNO indicator is closely related to the
maximal Lyapunov exponent, but it permits to determine  rapidly whether an
initial condition leads to a regular, quasi-periodic or a chaotic, irregular
solution. It is a very efficient tool that helps to detect orbital
resonances, their structure, and unstable regions in the phase space. 
Looking at a neighborhood of the analyzed initial conditions
is a profitable way of resolving the question whether the system dynamics are
robust to the fit errors.

The MEGNO tests reveal that both the LM1 and LM2 fits are related to
quasi-periodic motions of the HD12661 system.  The MEGNO signature for
the LM1 fit is shown in Fig.~{\ref{fig:fig4}c and a perfect
  convergence of the temporal value of MEGNO to~2 indicates a
  quasi-periodic evolution of the planetary system.  The MEGNO scan in
  the neighborhood of this fit, in the
  ($a_{\idm{c}},e_{\idm{c}})$-plane, is shown in the left panel of
  Fig.~\ref{fig:fig3}.  Actually, both fits are located in a
  relatively extended stable zone, in a proximity of the the 6:1 MMR.
  The 6:1 MMR is separated about $0.2$~au from the 11:2 MMR, which has
  been considered, up to now, as the closest MMR neighboring the
  system in the phase space. The evolution of orbital elements in the
  LM1 fit is shown in Fig.~\ref{fig:fig4}a,b.

The stable zone of the 6:1 MMR  is relatively very narrow in the range of
small $e_{\idm{c}}$. To illustrate its effect on the motion of the HD~12661
system we changed $a_{\idm{c}}$ in the LM2 fit, from the nominal value
$\simeq 2.78$~au to $2.745$~au, still keeping it in the error bound. The
MEGNO scan, in the ($a_{\idm{c}},e_{\idm{c}})$-plane, for such modified
initial condition is shown in the right panel of Fig.~\ref{fig:fig3}, and the
Jacobi elements are shown in Fig.~\ref{fig:fig4}d,e,f. Note that  in both
cases the SAR with the apsidal lines antialigned in the exact resonance is
present, so the system still remains in the large libration island found by
\cite{Lee2003}.

The scan in the $(i,\Delta\Omega)$-plane,
where $\Delta\Omega=\Omega_{\idm{c}}-\Omega_{\idm{b}}$,  
is shown in  Fig.~\ref{fig:fig5}, 
and it makes is possible to estimate the bounds on
masses in the systems with the same $i$, in terms of regular and chaotic
motions. 
For coplanar systems, the stable
zone extends up to $i \simeq 25^{\circ}$, but also two islands about $i
\simeq 10^{\circ}, 18^{\circ}$ exist. Whether the system can survive during
evolutionary time scales, in the chaotic regions, can be verified, in
general, only by long-term integrations.

\section*{Conclusions} 
The study of the RV observations  of the HD~12661 star revealed fits which
lead to significantly different dynamics when compared to the original
solution published by \cite{Fischer2003}. Combined genetic and gradient
methods of optimization helped us to find a global minimum of $\chi^2$ for
the RV model that incorporates the mutual interactions between planets. A
dynamical analysis shows that the new fits are related to a system close to
the 6:1 MMR, and locked into a deep secular resonance with apsidal lines
antialigned. The fits  preclude the proximity of the  11:2 MMR, which
has been suggested by the previously determined initial conditions. In the
phase space, the system lies in a
relatively  extended zone of stable, quasi-periodic
motions.  The closeness of the HD~12661 system to the low-order 6:1 MMR 
adds a new inquiring case to the 3:1 MMR in the 55~Cnc system, the 7:3
MMR in the 47~UMa system, and the 5:2 MMR of Jupiter and Saturn. It is
difficult to claim that the HD~12661 system is locked exactly in the 6:1
MMR, but our results suggest that very likely it is close to this resonance.
Hopefully, the observational window of the HD~12661 will cover soon 
two orbital periods of the outer companion, and the updated 
set of measurements will help to refine the fits again.

The analysis of the RV data benefits from a {\em global} approach, similarly
to the stability studies. For  planetary systems with large, and not very
distant planets, the effects of mutual interactions and a proper dynamical
interpretation of the orbital fits are vital for understanding the dynamics
and for resolving all the information hidden in the observations. 
Their analysis requires much care, because
omitting even  subtle factors can lead to quite a different
interpretation of the same data set. 

\section*{Acknowledgments}
We want to thank Dr. Gregory Laughlin and the anonymous Referee for
invaluable remarks and comments that improved the manuscript.
We are indebted to Zbroja for correcting the manuscript. Calculations in this
paper were performed on the HYDRA computer-cluster,  supported by the
Polish Committee for Scientific Research, Grants No.~5P03D~006~20 and
No.~2P03D~001~22. This work is supported by the Polish Committee for
Scientific Research, Grant No.~2P03D~001~22.
\bibliographystyle{apj}
\bibliography{ms}

\newpage
%
%
\begin{table}[th]
\caption{Osculating, astrocentric Keplerian
elements of the HD~12661 planetary system for the epoch
of the first observation (JD=2450831.608). 
Mass of the host star is 1.07~$M_{\sun}$.
}
\vskip 0.5cm
\centering
 \begin{tabular}{lcccc}
\hline
Fit &\multicolumn{2}{c}{LM1 ($N_p$=10+2)} 
  &\multicolumn{2}{c}{Mean (LM2) ($N_p$=10+2) } \\
\hline
Parameter \ \ \ \ \ \ \  & Planet b & Planet c & Planet b & Planet c \\
\hline
$\mbox{m}_{\idm{pl}} \sin i$ [M$_{\mathrm{J}}$] \dotfill
& 2.33 (0.04)  &  1.69 (0.11)
& 2.32 (0.04)  &  1.63 (0.11)  \\
$a$ [au] \dotfill  
                &  0.822 (0.001) &  2.804 (0.11) &
                   0.823 (0.001) &  2.781 (0.11) \\

$e$ \dotfill  &
                   0.349 (0.016)    &  0.084 (0.054)                 
                 & 0.343 (0.016)      & 0.128 (0.054)   \\
$\omega$ [deg]\dotfill 
                       & 115.2 (3.9) & 292.8 (34.1)              
                       & 113.8 (3.9) & 303.1 (34.1)  \\
$M$ [deg]\dotfill 
                    & 129.4 (4.7) & 354.3 (42.9)                
                    & 132.1 (4.7) & 342.5 (42.9)\\
$V_0$ (Lick) [ms$^{-1}$] \dotfill
        &     \multicolumn{2}{c}{-7.3 (1.9)} &              
               \multicolumn{2}{c}{-6.3 (1.9)} \\
$V_1$ (Keck) [ms$^{-1}$] \dotfill 
        &     \multicolumn{2}{c}{-12.7 (1.9)} &              
               \multicolumn{2}{c}{-12.2 (1.9)} \\
$\sqrt{\chi^2_{\nu}}$ \dotfill 
         &    \multicolumn{2}{c}{1.37} &              
               \multicolumn{2}{c}{0.93} \\
RMS [ms$^{-1}$]  \dotfill 
         &    \multicolumn{2}{c}{7.88} &              
               \multicolumn{2}{c}{7.44} \\
\hline
\end{tabular} 
\label{tab:fits}
\end{table}

\newpage
\eject
%
%
\begin{figure}[th]
\centering
\hbox{ \includegraphics[]{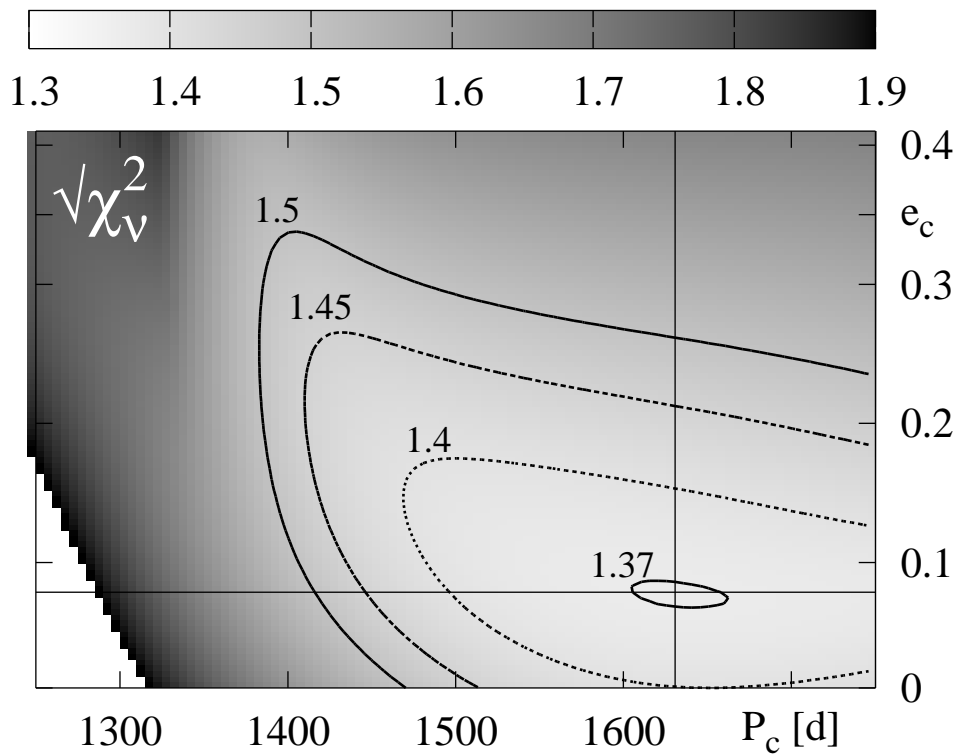}}
\caption{
The minimum values of $\sqrt{\chi^2_{\nu}}$
as a function of $(P_{\idm{c}},
e_{\idm{c}})$ for the  2-Keplerian, Jacobi 
model of the RV observations ($N_p=10+2$). 
The global 
minimum is marked by the intersection of two lines,
corresponding to $P_{\idm{c}} \simeq 1630$~d, $e_{\idm{c}}\simeq 0.079$.
See the text for an explanation.
}
\label{fig:fig1}
\end{figure}
\newpage
\eject

%
%
\begin{figure}[th]
\centering
\hbox{ \includegraphics[]{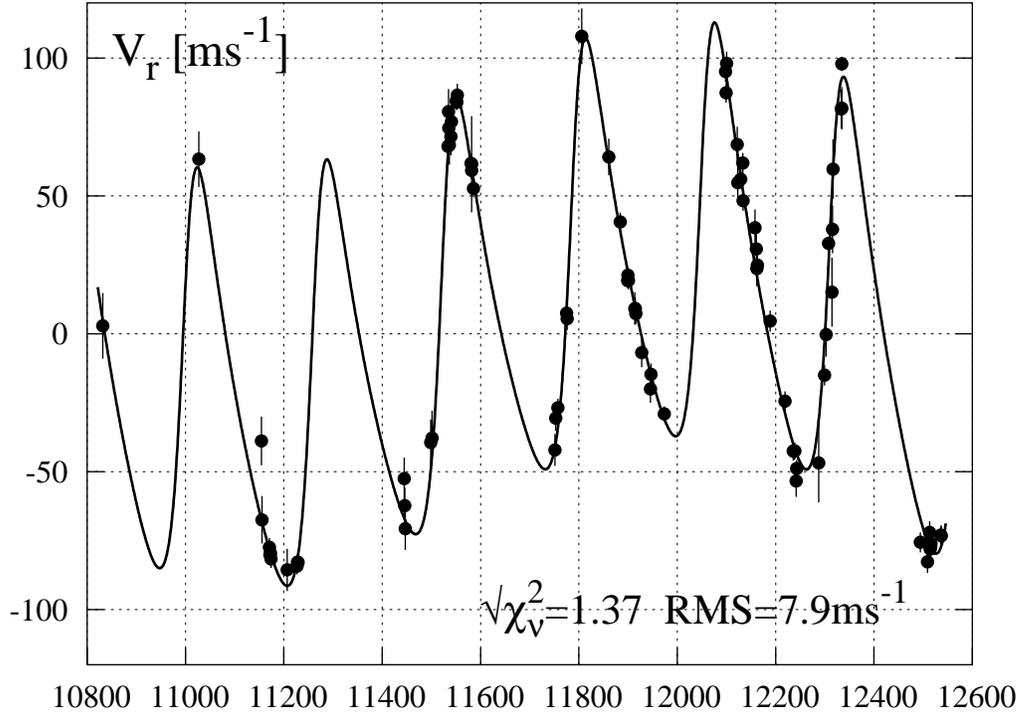}}
\caption{
The synthetic radial velocity curve defined by the LM1 fit (see Table~1).
Time is given in JD. Filled circles mark the observations published
by \cite{Fischer2003}.
}
\label{fig:fig2}
\end{figure}
\newpage
\eject

%
%
\begin{figure*}[th]
\hspace*{-1cm}
\hbox{
          \includegraphics[width=9cm]{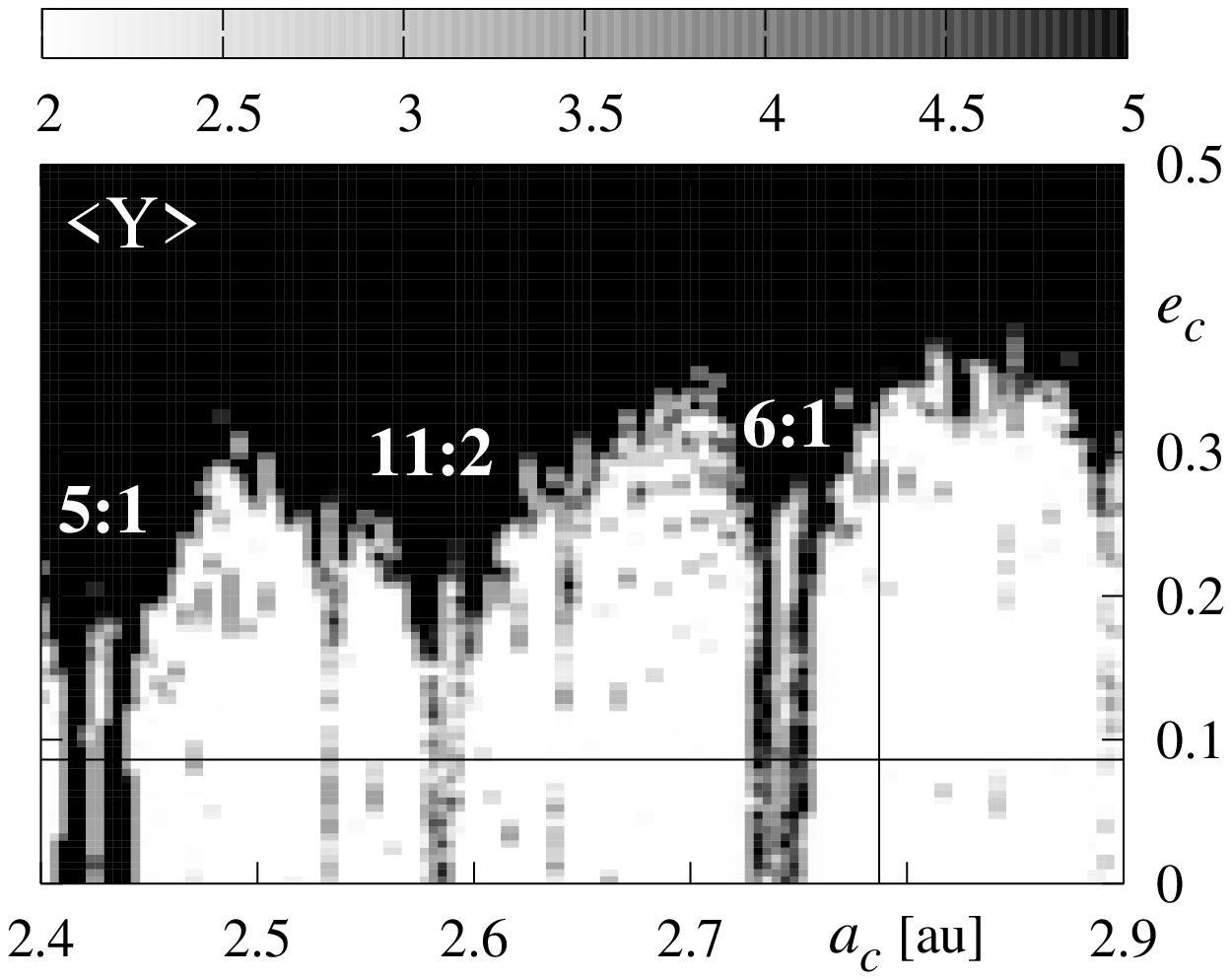}
          \includegraphics[width=9cm]{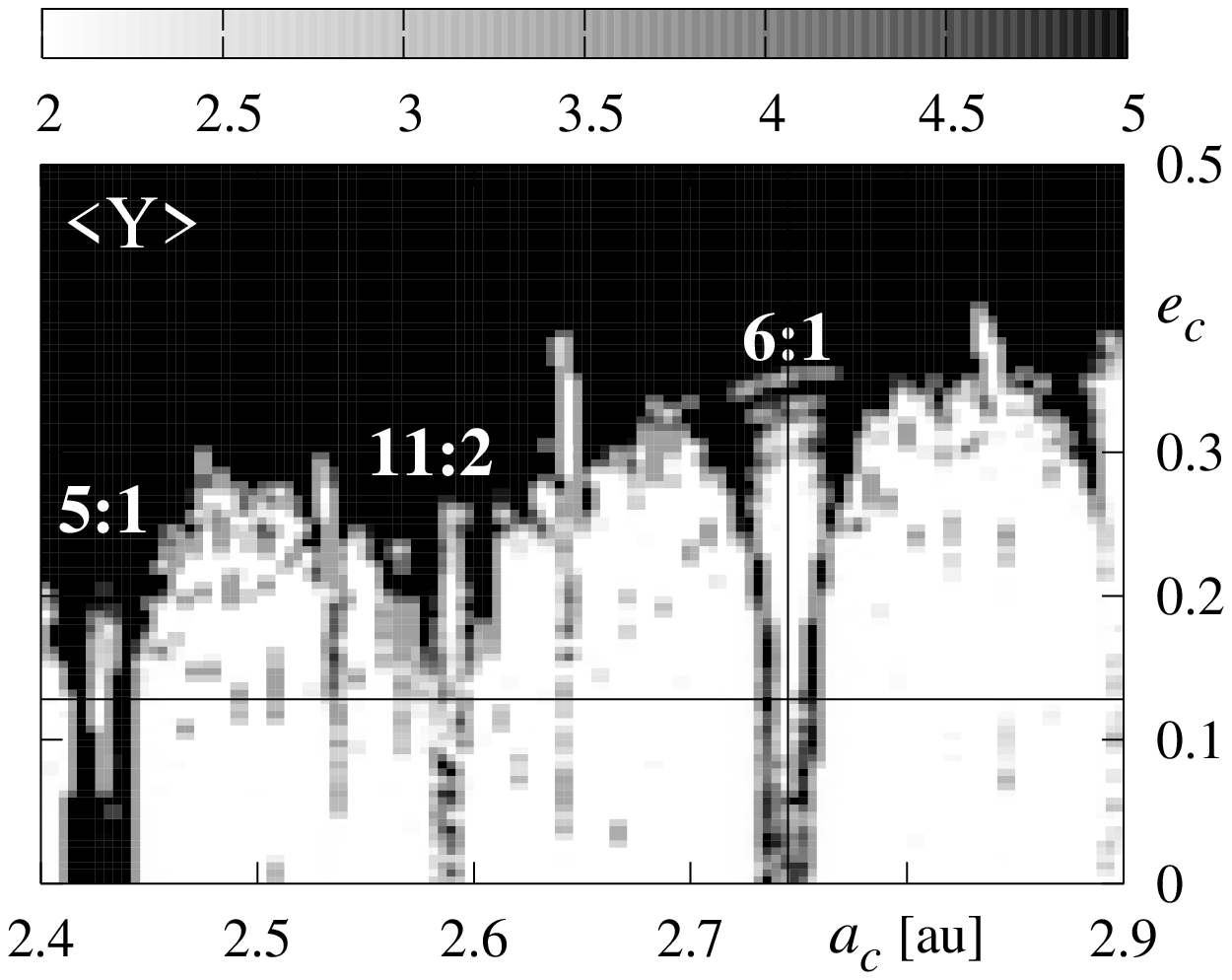}
}
\caption{
The MEGNO maps in the $(P_{\idm{c}},e_{\idm{c}})$-plane  for the initial
elements found in this paper. The left panel is for the LM1 fit, the right
panel is for the modified LM2 fit
(as explained in the text). Zones centered about $\simeq 2$  correspond to
regular, quasi-periodic motions of the  HD~12661 system. The intersection of
two lines marks the fitted parameters, the MMRs relevant to
our discussion are labeled. The resolution of the scan is $120 \times 100$
data points.
}
\label{fig:fig3}
\end{figure*}
\newpage

%
%
%
\begin{figure}[th]
\centering
\hbox{ \includegraphics[]{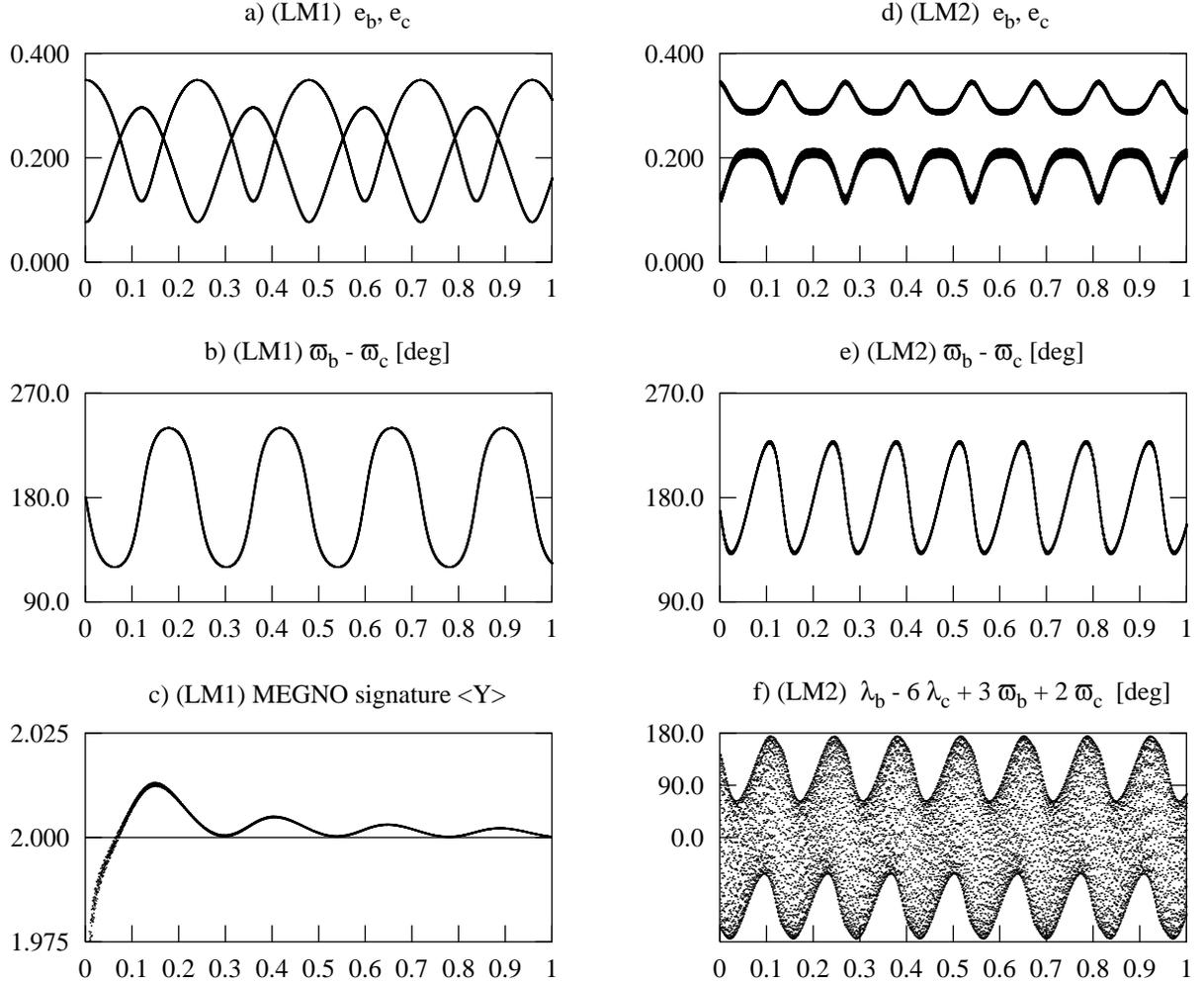}}
\caption
{ 
The evolution of Keplerian orbital elements 
related to Jacobi coordinates for the initial conditions
found in this paper. The left column is for the LM1 fit, the right column is
for the modified  LM2 fit. Note the presence of the secular apsidal resonance---its 
critical argument $\varpi_{\idm{b}}-\varpi_{\idm{c}}$ librates about
$180^{\circ}$ (panels b,e). The 
modified LM2 fit leads to the 6:1 MMR (panel f), while
for the LM1 fit the same critical argument circulates. Note the stabilizing
effect of the 6:1 MMR on the eccentricities of the planets (panels a,d). 
Panel c is for MEGNO signature of the LM1 fit and it indicates
a quasi-periodic, stable  motion of the planetary system.
Time is given in $10^5$~yr.
}
\label{fig:fig4}
\end{figure}
\newpage
\eject

%
%
\begin{figure}[th]
\centering
\hbox{ \includegraphics[]{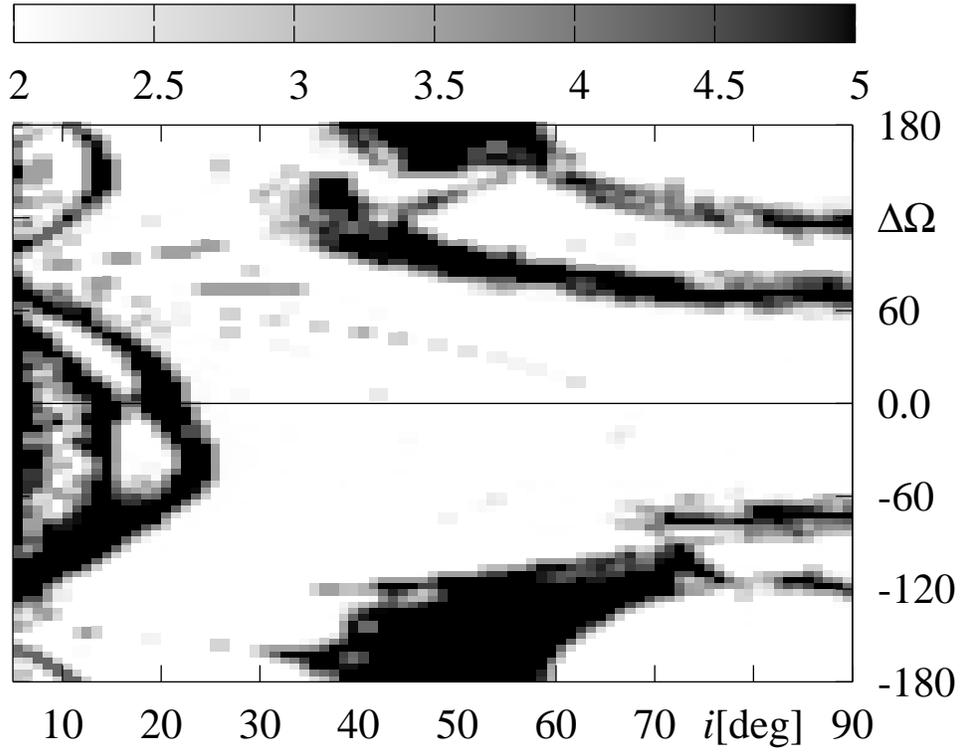} }
\caption{
Stability regions in the plane of the system inclination $i$ vs  the relative
inclination of orbits (expressed through the difference of the longitudes of
nodes, $\Delta\Omega=\Omega_{\idm{c}}-\Omega_{\idm{b}}$), 
derived for the LM2 fit. The resolution of the scan is $85 \times 90$
data points.
}
\label{fig:fig5}
\end{figure}
\newpage
\eject

\end{document}